\documentclass[11pt]{article}
\begin{document}
\title{
Negative Volterra Flows and Mixed Volterra Flows and Their
Infinitely Many Conservation Laws}
\author{
Zuo-nong Zhu $^{1}$ and Hon-Wah Tam $^{2}$\\
1. Department of Mathematics, Shanghai Jiao Tong University,\\
Shanghai, 200030, P.R. China\\
2. Department of Computer Science, Hong Kong Baptist University,\\
224 Waterloo Road, Kowloon, Hong Kong, P. R. China\\
\footnote{E-mail: znzhu@online.sh.cn}}
\date{}
\maketitle
\begin{abstract}
In this article, by means of  considering an isospectral operator equation which corresponds to the Volterra lattice, 
and constructing opportune time evolution problems with negative powers of spectral parameter, and using discrete zero curvature 
representation, negative Volterra flows are proposed. We also propose the mixed Volterra flows, which come from positive and negative volterra flows. From the Lax representation,
we demonstrate the existence of infinitely many conservation laws for the two flows
and give the corresponding conserved densities and the associated fluxes
formulaically. Thus their integrability is further confirmed.
\end{abstract}
\setcounter{section}{0}
\setcounter{equation}{0}
\section{Introduction}
Nonlinear integrable lattice systems have been received considerable
attention in recent years. It is well known that discrete lattice systems not only have 
rich mathematical structures but also have many applications in science, such as mathematical physics, numerical analysis, computer science, statistical physics, quantum physics, and so on. Among the most famous and well studied integrable lattice, the Volterra lattice
\begin{eqnarray}
\dot{u}_n=u_n(u_{n+1}-u_{n-1})
\end{eqnarray}
is one of the popular models. The Volterra lattice (1.1) has been studied extensively [1-6]. Its Lax pair is presented by
\begin{eqnarray}
E\psi_n=U_n\psi_n,\nonumber\\
\frac{d\psi_n}{dt}=V_n\psi_n,
\end{eqnarray}
where
\begin{eqnarray}
U_n=\left(
\begin{array}{cc}
\lambda&u_n\\
-1&0
\end{array}\right),
\qquad
V_n=\left(
\begin{array}{cc}
u_n&\lambda u_n\\
-\lambda &u_{n-1}-\lambda^2
\end{array}\right),
\end{eqnarray}
Recently, by using zero curvature equation,
\begin{eqnarray}
\dot{U}_n=V_{n+1}U_n-U_nV_n,
\end{eqnarray}
and constructing time evolution matrix $V_n$ with negative powers of spectral parameter $\lambda$, Pritula and Vekslerchik [7] proposed the following negative Volterra flows:
\begin{eqnarray}
\tau_{n-1}\tau_{n+1}\frac{\partial}{\partial t_{j+1}}ln\frac{\tau_{n+1}}{\tau_{n-1}}+\tau_n^2\frac{\partial^2}{\partial t_1\partial t_j}ln \tau_n=0,\qquad j=1,2,....
\end{eqnarray}
and 
\begin{eqnarray}
\tau_{n-1}\tau_{n+1}\frac{\partial}{\partial t_{1}}ln\frac{\tau_{n+1}}{\tau_{n-1}}=\tau_n^2.
\end{eqnarray}
Here $u_n$ is presented by tau-function
\begin{eqnarray}
u_n=\frac{\tau_{n+1}\tau_{n-2}}{\tau_n\tau_{n-1}}
\end{eqnarray}
Dark-soliton solutions of negative Volterra flows (1.5) and (1.6) are also given [6].
The negative Volterra lattice hierarchy (1.5)-(1.6) is different from the most known integrable lattice hierarchy. We note
field-function $u_n$ is dependent on discrete space variable $n$ and continuum time variable $t$.
However, in lattice hierarchy (1.5)-(1.6), the property of field function $u_n$ at $t_{j+1}$ is dependent on its property at $t_1$ and $t_{j}$. In fact, evolution equation at different time $t_j$ should be independent.
In this paper, motived by the idea proposed in [7], that is by means of constructing proper time evolution matrix $V_n$ with negative powers of spectral parameter $\lambda$, we derive another negative Volterra flows. We also obtain a mixed Volterra flows which come from positive Volterra flows and negative Volterra flows. It is well known that the existence of infinitely many conservation laws is very important indicator of integrability of the system. From physical view and numerical analysis, it is also very useful to know whether exist
conservation laws for a lattice system. Using the explicit matrix Lax representation and following the method studied in [8-12], we demonstrate the existence of infinitely many conservation laws for the negative Volterra flows and mixed Volterra flows  
and also give the corresponding conserved densities and the associated fluxes
formulaically. 
\setcounter{equation}{0}
\section{Negative Volterra flows related to isospectral problem (1.2)}
In order to derive the negative Volterra flows from discrete zero curvature representation (1.4), we should construct opportune time evolution equation,
\begin{eqnarray}
\frac{d\psi_n}{dt}=V_n^{(m)}\psi_n.
\end{eqnarray}
Set
\begin{eqnarray}
V_n^{(m)}=\left(
\begin{array}{cc}
A^{(m)}(\lambda)& B^{(m)}(\lambda)\\
C^{(m)}(\lambda)& D^{(m)}(\lambda)
\end{array}\right)
\end{eqnarray}
It is easy to get the following equations:
\begin{eqnarray}
B^{(m)}=-u_nEC^{(m)},\qquad
D^{(m)}=E^{-1}A^{(m)}+\lambda C^{(m)}
\end{eqnarray}
and
\begin{eqnarray}
\lambda (E-1)A^{(m)}+u_{n+1}E^2C^{(m)}-u_nC^{(m)}=0\\
\dot{u}_n=u_n[(E-E^{-1})A^{(m)}+\lambda(E-1)C^{(m)}]
\end{eqnarray}
Let
\begin{eqnarray}
A^{(m)}(\lambda)=\sum_{j=1}^{m}a_{m-j}\lambda^{-2j},\qquad C^{(m)}(\lambda)=\sum_{j=1}^{m}c_{m-j}\lambda^{-2j+1}.
\end{eqnarray}
From discrete zero curvature equation (1.4), $a_j, c_j (j=0,1,....,m-1)$ must satisfy the 
following equations:
\begin{eqnarray}
(E-E^{-1})a_{j}+(E-1)c_{j-1}=0,\qquad j=1,2,...,m-1\\
(E-1)a_j+u_{n+1}E^2c_j-u_nc_j=0, \qquad j=1,2,.....,m-1\\
(E-E^{-1})a_0=0,\qquad
u_{n+1}E^2c_0-u_nc_0=0
\end{eqnarray}
and we obtain the negative Volterra flows,
\begin{eqnarray}
\dot{u}_n=u_n(E-1)c_{m-1}, \qquad m\geq 1.
\end{eqnarray}
When field function $u_n$ is presented by tau-function (1.7), equation (2.10) is written as
\begin{eqnarray}
\frac{\partial}{\partial t}ln\frac{\tau_{n+1}}{\tau_{n-1}}=Ec_{m-1}, \qquad m\geq 1.
\end{eqnarray}
How to determine $c_j(j=0,1,2,.....)$? First we choose $a_0=0$ or $a_0=1$,
and we have
\begin{eqnarray}
c_0=\frac{\tau_{n-1}^2}{\tau_{n-2}\tau_n}.
\end{eqnarray}
Then we can determine $a_j$ and $c_j$ (j=1,2,....,m-1) from equations (2.7)-(2.8) via the following path:
$$c_0\rightarrow a_1\rightarrow c_{1} \rightarrow a_2\rightarrow c_2
\rightarrow .........\rightarrow a_{m-1}\rightarrow c_{m-1}$$ 
Note that the general solutions of the difference equations
\begin{eqnarray}
(E^{-1}+1)X(n)=F(n),\\
(E^2-1)Y(n)=G(n),
\end{eqnarray}
can be written as
\begin{eqnarray}
X(n)=c(-1)^n+(E^{-1}+1)^{-1}F(n)=c(-1)^n+\sum_{k=0}^{\infty}(-1)^kE^{-k}F(n),\\
Y(n)=c+d(-1)^n+(E^2-1)^{-1}G(n)=c+d(-1)^n-\sum_{k=0}^{\infty}E^{2k}G(n),
\end{eqnarray}
where $c$ and $d$ are two arbitrary constants. Solving equations (2.7)-(2.8), we have the following results:
\begin{eqnarray}
a_1=-(1+E^{-1})^{-1}c_0=\sum_{k=0}^{\infty}(-1)^{k+1}E^{-k}\frac{\tau_{n-1}^2}{\tau_n\tau_{n-2}},\\
c_1=\frac{\tau_{n-1}^2}{\tau_{n-2}\tau_n}[c+d(-1)^n-\sum_{k=0}^{\infty}E^{2k}(\frac{\tau_n^4}{\tau_{n-1}^2\tau_{n+1}^2}+\frac{2\tau_{n}^2}{\tau_{n-1}\tau_{n+1}}\sum_{j=0}^{\infty}(-1)^{j+1}E^{-j}\frac{\tau_{n-1}^2}{\tau_{n-2}\tau_n})],\\
a_j=\sum_{k=0}^{\infty}(-1)^{k+1}E^{-k}c_{j-1}, \qquad j=2,3,.....,m-1\\
c_j=\frac{\tau_{n-1}^2}{\tau_{n-2}\tau_n}(E^2-1)^{-1}[\frac{\tau_{n}^2}{\tau_{n-1}\tau_{n+1}}(1-E)a_j], \qquad j=2,3,.....,m-1
\end{eqnarray}
The first negative Volterra flow is 
\begin{eqnarray}
\frac{\partial}{\partial t}ln\frac{\tau_{n+1}}{\tau_{n-1}}=\frac{\tau_{n}^2}{\tau_{n+1}\tau_{n-1}}
\end{eqnarray}
and the second negative Volterra flow is
\begin{eqnarray}
\frac{\partial}{\partial t}ln\frac{\tau_{n+1}}{\tau_{n-1}}=Ec_1,
\end{eqnarray}
where $c_1$ is presented by equation (2.18).
\setcounter{equation}{0}
\section{Mixed Volterra flows related to isospectral problem (1.2)}
It is well known that positive Volterra flows related to isospectral problem (1.2) can be obtained by the following approach.
Set
\begin{eqnarray}
V_n^{(s)}=\left(
\begin{array}{cc}
G^{(s)}(\lambda)& -u_nEH^{(s)}(\lambda)\\
H^{(s)}(\lambda)& E^{-1}G^{(s)}+\lambda H^{(s)}
\end{array}\right)
\end{eqnarray}
where
\begin{eqnarray}
G^{(s)}(\lambda)=\sum_{j=0}^{s}g_{s-j}\lambda^{2j},\qquad H^{(s)}(\lambda)=\sum_{j=0}^{s}h_{s-j}\lambda^{2j+1},
\end{eqnarray}
and  $g_j, h_j (j=0,1,....,s)$ are determined by the following equations:
\begin{eqnarray}
(E-E^{-1})g_{j}+(E-1)h_{j+1}=0,\qquad j=0,1,2,...,s-1\\
(E-1)g_j+u_{n+1}E^2h_j-u_nh_j=0, \qquad j=0,1,2,.....,s\\
(E-1)h_0=0,
\end{eqnarray}
then positive Volterra flows are proposed,
\begin{eqnarray}
\dot{u}_n=u_n(E-E^{-1})g_s,\qquad s\geq 0
\end{eqnarray}
Let $s=0$, equation (3.6) reduces to the Volterra lattice (1.1). The second positive Volterra flow corresponding 
to $s=1$ is written as
\begin{eqnarray}
\dot{u}_n=u_nu_{n+1}(u_n+u_{n+1}+u_{n+2})-u_nu_{n-1}(u_n+u_{n-1}+u_{n-2})
\end{eqnarray}
Mixing positive and negative Volterra flows (2.10) and (3.6), we obtain the so-called mixed Volterra flows
\begin{eqnarray}
\dot{u}_n=u_n[(E-1)c_{m-1}+(E-E^{-1})g_s], \qquad m\geq 1, s\geq 0 
\end{eqnarray}
where $u_n$ is presented by tau-function (1.7). It is obvious that
mixed Volterra flows admit the matrix Lax pairs with $U_n$ and $V_n^{(m,s)}$, where $V_n^{(m,s)}$ possesses form
\begin{eqnarray}
V_n^{(m,s)}=\left(
\begin{array}{cc}
A^{(m)}+G^{(s)}&-u_nE(C^{(m)}+H^{(s)})\\
C^{(m)}+H^{(s)}&E^{-1}(A^{(m)}+G^{(s)})+\lambda(C^{(m)}+H^{(s)})
\end{array}\right)
\end{eqnarray}
Mixed Volterra flows (3.8) can be written in the form:
\begin{eqnarray}
\frac{\partial}{\partial t}ln\frac{\tau_{n+1}}{\tau_{n-1}}=Ec_{m-1}+(E+1)g_s,\qquad m\geq 1, s\geq 0 
\end{eqnarray}
Set $m=1, s=0$, we obtain a mixed Volterra lattice equation
\begin{eqnarray}
\frac{\partial}{\partial t}ln\frac{\tau_{n+1}}{\tau_{n-1}}=\frac{\tau_n^2}{\tau_{n-1}\tau_{n+1}}
+\frac{\tau_{n+1}\tau_{n-2}}{\tau_{n-1}\tau_{n}}
+\frac{\tau_{n-1}\tau_{n+2}}{\tau_{n}\tau_{n+1}}
\end{eqnarray}
Set $m=1, s=1$, another mixed Volterra lattice equation is given,
\begin{eqnarray}
\frac{\partial}{\partial t}ln\frac{\tau_{n+1}}{\tau_{n-1}}=\frac{\tau_n^2}{\tau_{n-1}\tau_{n+1}}
+\frac{\tau_{n+1}\tau_{n-2}}{\tau_{n-1}\tau_{n}}(\frac{\tau_{n}\tau_{n-3}}{\tau_{n-2}\tau_{n-1}}
+\frac{\tau_{n+1}\tau_{n-2}}{\tau_{n-1}\tau_{n}}+\frac{\tau_{n+2}\tau_{n-1}}{\tau_{n}\tau_{n+1}})\nonumber\\
+\frac{\tau_{n-1}\tau_{n+2}}{\tau_{n}\tau_{n+1}}(\frac{\tau_{n-2}\tau_{n+1}}{\tau_{n}\tau_{n-1}}+
\frac{\tau_{n-1}\tau_{n+2}}{\tau_{n}\tau_{n+1}}+\frac{\tau_{n}\tau_{n+3}}{\tau_{n+1}\tau_{n+2}})
\end{eqnarray}
\setcounter{equation}{0}
\section{Infinitely many conservation laws for negative Volterra flows (2.11) and mixed Volterra flows (3.10)}
For a lattice equation
\begin{eqnarray}
F(\dot{q}_n,\ddot{q}_n,...,q_{n-1},q_n,q_{n+1},...)=0,
\end{eqnarray}
if there exist functions $\rho_n$ and $J_n$, such that
\begin{equation}
\dot{\rho}_n|_{F=0}=J_{n}-J_{n+1},
\end{equation}
then equation (4.2) is called the conservation law of equation (4.1),
where $\rho_n$ is
the conserved density and $J_n$ is the associated flux. Suppose equation
(4.1) has conservation law (4.2) and $J_n$ is bounded for all $n$ and vanishes at the boundaries, then $\sum_n\rho_n=c$ with $c$ being arbitrary constant is an integral of motion of lattice equation (4.1). 
In this section, we first demonstrate the existence of infinitely many conservation laws for lattice hierarchy related to isospectral problem (1.2) by means of the  explicit matrix Lax representation, and then we derive infinitely many conservation laws for negative Volterra lattice hierarchy
and mixed Volterra lattice hierarchy in details and give the corresponding conserved densities and the associated fluxes formulaically.\\
{\underline{4.1 Infinitely many conservation laws for lattice hierarchy associated with  isospectral problem (1.2)}}\\
It is obvious that isospectral problem (1.2) is equivalent to
\begin{eqnarray}
\psi_{2,n+1}=\lambda\psi_{2,n}-u_{n-1}\psi_{2,n-1}.
\end{eqnarray}
Let $\Gamma_n=\frac{\psi_{2,n-1}}{\psi_{2,n}}$ and note that
\begin{eqnarray}
\frac{(\psi_{2,n+1}\psi_{2,n}^{-1})_t}{\psi_{2,n+1}\psi_{2,n}^{-1}}
=\frac{(\psi_{2,n+1})_t}{\psi_{2,n+1}}-\frac{(\psi_{2,n})_t}{\psi_{2,n}},
\end{eqnarray}
then we obtain
\begin{eqnarray}
\frac{\partial}{\partial t}
[\ln (\lambda -u_{n-1}\Gamma_{n})]
=Q_{n+1}-Q_n,
\end{eqnarray}
where
\begin{eqnarray}
Q_n=\frac{(\psi_{2,n})_t}{\psi_{2,n}}=V_{21}^{(m)}(u_{n-1}\Gamma_{n}-\lambda)
+V_{22}^{(m)}.
\end{eqnarray}
It follows from (4.3) that
\begin{equation}
u_{n-1}\Gamma_n\Gamma_{n+1}-\lambda\Gamma_{n+1}+1=0.
\end{equation}
(4.7) is a discrete Ricatti type equation, which can be given a series solution.
Suppose the eigenfunction $\psi_2(n,t,\lambda)$
is the analytical function of the arguments and expand $\Gamma_n$ with respect to $\lambda$ by the Taylor series
\begin{eqnarray}
\Gamma_n=\sum_{j=1}^{\infty}\lambda^{-j}w_n^{(j)},
\end{eqnarray}
and substituting eq. (4.8) into eq. (4.7), we obtain $w_n^{(j)}$ recursively,
\begin{eqnarray}
w_n^{(1)}=1,\qquad
w_n^{(2j)}=0, \qquad 
w_n^{(2j+1)}=u_{n-2}\sum_{l+s=2j}w_{n-1}^{(l)}w_n^{(s)}, \qquad j=1,2,3,.....
\end{eqnarray}
It follows that
\begin{eqnarray}
w_n^{(3)}=u_{n-2}, \qquad w_{n}^{(5)}=u_{n-2}(u_{n-2}+u_{n-3}),
\end{eqnarray}
$$................$$
From eq. (4.5) we have
\begin{eqnarray}
\frac{\partial}{\partial t}\sum_{k=1}^{\infty}\frac{\Phi^k}{k}
=Q_{n}-Q_{n+1},
\end{eqnarray}
where
\begin{eqnarray}
\Phi=\lambda^{-1}u_{n-1}\Gamma_n=\sum_{j=1}^{\infty}\lambda^{-2j}u_{n-1}{w}_n^{(2j-1)}
\end{eqnarray}
It follows from eq. (4.11) that
\begin{eqnarray}
\frac{\partial }{\partial t} 
\sum_{j=1}^{\infty}\lambda^{-2j}\rho_n^{(j)}=Q_{n}-Q_{n+1},
\end{eqnarray}
where
\begin{eqnarray}
\rho_n^{(j)}=v_{2j-1}+\frac{1}{2}\sum_{l_1+l_2=2j-2}v_{l_1}v_{l_2}+\frac{1}{3}\sum_{l_1+l_2+l_3=2j-3}v_{l_1}v_{l_2}v_{l_3}
+.....+\nonumber\\
\frac{1}{j-2}\sum_{l_1+l_2+...+l_{j-2}=2j-j+2}v_{l_1}v_{l_2}....v_{l_{j-2}}+
v_1^{j-2}v_3+\frac{1}{j}v_1^{j}.
\end{eqnarray}
with
\begin{eqnarray}
v_j=u_{n-1}w_n^{(j)}
\end{eqnarray}
Making a comparison of the powers of $\lambda$ on both sides of eq. (4.13),
we obtain infinitely many conservation laws for lattice hierarchy related to isospectral problem (1.2),
\begin{eqnarray}
\rho_{n,t}^{(j)}=J_{n}^{(2j-1)}-J_{n+1}^{(2j-1)}, \qquad j=1,2,3,......
\end{eqnarray}
{\underline{4.2 Infinitely many conservation laws for negative Volterra flows and mixed Volterra flows}}\\  
For negative Volterra lattice hierarchy (2.11), note that
\begin{eqnarray}
Q_n=E^{-1}A^{(m)}+u_{n-1}C^{(m)}\Gamma_n,
\end{eqnarray}
we obtain its infinitely many conservation laws, where the associated fluxes $J_n^{(j)}$
are written as
\begin{eqnarray}
J_n^{(j)}=E^{-1}a_{m-j}+u_{n-1}\sum_{i=0}^{j-1}c_{m-j+i}w_n^{(2i+1)},\qquad
j=1,2,.....,m\nonumber\\
J_n^{(j)}=u_{n-1}\sum_{i=1}^{m}c_{m-i}w_n^{(2j-2i+1)}, j=m+1, m+2,.......
\end{eqnarray}
For mixed Volterra lattice hierarchy (3.10), note that
\begin{eqnarray}
Q_n=E^{-1}(A^{(m)}+G^{(s)})+u_{n-1}(C^{(m)}+H^{(s)})\Gamma_n.
\end{eqnarray}
Further we have 
\begin{eqnarray}
Q_n=\sum_{j=1}^{\infty}q_j\lambda^{-2j}
\end{eqnarray}
where
\begin{eqnarray}
q_j=E^{-1}a_{m-j}+u_{n-1}(\sum_{i=0}^{j-1}c_{m-j+i}w_n^{(2i+1)}+\sum_{i=0}^{s}h_{s-i}w_n^{(2j+2i+1)}), \qquad j=1,2,....,m\nonumber\\
q_{m+j}=u_{n-1}(\sum_{i=1}^{m}c_{m-i}w_n^{(2m+2j-2i+1)}+\sum_{i=0}^{s}h_{s-i}w_n^{(2m+2j+2i+1)}),\qquad j=1,2,3,....
\end{eqnarray}
We thus obtain infinitely many conservation laws for mixed Volterra lattice hierarchy, where the associated fluxes $J_n^{(j)}$ are presented by $q_j$.\\
{\bf Example 1}.
For the first negative Volterra flow (2.21), continuous time evolution equation is
\begin{eqnarray}
\frac{d\psi_n(\lambda)}{dt}=V_n^{(1)}\psi_n(\lambda), \qquad V_n^{(1)}=\left(
\begin{array}{cc}
0&-\frac{\tau_{n-2}\tau_n}{\lambda \tau_{n-1}^2}\\
\frac{\tau_{n-1}^2}{\lambda \tau_{n-2}\tau_n}&\frac{\tau_{n-1}^2}{\tau_{n-2}\tau_n}
\end{array}\right)
\end{eqnarray}
Note that
\begin{eqnarray}
Q_n=\frac{\tau_{n-1}^2}{\lambda \tau_{n-2}\tau_n}(\frac{\tau_{n}\tau_{n-3}}{\tau_{n-1}\tau_{n-2}}\Gamma_n-\lambda)+\frac{\tau_{n-1}^2}{\tau_{n-2}\tau_n}
=\sum_{j=1}^{\infty}J_n^{(2j-1)}\lambda^{-2j},
\end{eqnarray}
where
\begin{eqnarray}
J_n^{(2j-1)}=\frac{\tau_{n-1}\tau_{n-3}}{\tau_{n-2}^2}w_n^{(2j-1)},\qquad j=1,2,.....
\end{eqnarray}
So, the conserved densities $\rho_n^{(j)}$(j=1,2,3,....) and the associated flux $J_n^{(2j-1)}$ (j=1,2,....) for flow (2.21) are given, where $J_n^{(2j-1)}$ is presented by equation (4.24).\\
{\bf Example 2}.
For the mixed Volterra lattice equation (3.11),  continuous time evolution equation is
\begin{eqnarray}
\frac{d\psi_n(\lambda)}{dt}=V_n^{(1,0)}\psi_n(\lambda), \qquad V_n^{(1,0)}=\left(
\begin{array}{cc}
\frac{\tau_{n+1}\tau_{n-2}}{\tau_{n}\tau_{n-1}}&\frac{\lambda\tau_{n+1}\tau_{n-2}}{\tau_{n}\tau_{n-1}}-\frac{\tau_{n-2}\tau_n}{\lambda \tau_{n-1}^2}\\
\frac{\tau_{n-1}^2}{\lambda \tau_{n-2}\tau_n}-\lambda&\frac{\tau_{n}\tau_{n-3}}{\tau_{n-1}\tau_{n-2}}+\frac{\tau_{n-1}^2}{\tau_{n-2}\tau_n}-\lambda^2
\end{array}\right)
\end{eqnarray}
Note that
\begin{eqnarray}
Q_n=(\frac{\tau_{n-1}^2}{\lambda \tau_{n-2}\tau_n}-\lambda)(\frac{\tau_{n}\tau_{n-3}}{\tau_{n-1}\tau_{n-2}}\Gamma_n-\lambda)+\frac{\tau_{n}\tau_{n-3}}{\tau_{n-1}\tau_{n-2}}+\frac{\tau_{n-1}^2}{\tau_{n-2}\tau_n}-\lambda^2
=\sum_{j=1}^{\infty}J_n^{(2j-1)}\lambda^{-2j},
\end{eqnarray}
where
\begin{eqnarray}
J_n^{(2j-1)}=\frac{\tau_{n-1}\tau_{n-3}}{\tau_{n-2}^2}w_n^{(2j-1)}-\frac{\tau_{n}\tau_{n-3}}{\tau_{n-2}\tau_{n-1}}w_n^{(2j+1)},\qquad j=1,2,.....
\end{eqnarray}
Therefore, the conserved densities $\rho_n^{(j)}$ and the associated flux $J_n^{(2j-1)}$ (j=1,2,....) for eq. (3.11) are given, where $J_n^{(2j-1)}$ is presented by equation (4.27).\\
{\bf Example 3}.
For the second negative Volterra flow (2.22), the associated continuous time evolution equation is written as
\begin{eqnarray}
\frac{d\psi_n(\lambda)}{dt}=V_n^{(2)}\psi_n(\lambda), \qquad V_n^{(2)}=\left(
\begin{array}{cc}
\frac{a_1}{\lambda^2}&-\frac{\tau_{n+1}\tau_{n-2}}{\tau_{n}\tau_{n-1}}(\frac{Ec_1}{\lambda}+\frac{\tau_{n}^2}{\lambda^3\tau_{n-1}\tau_{n+1}})\\
\frac{c_1}{\lambda}+\frac{\tau_{n-1}^2}{\lambda^3\tau_{n-2}\tau_{n}}
&c_1+\frac{E^{-1}a_1}{\lambda^2}+\frac{\tau_{n-1}^2}{\lambda^2\tau_{n-2}\tau_{n}}
\end{array}\right)
\end{eqnarray}
where $a_1$ and $c_1$ are presented by equations (2.17) and (2.18).
Note that
\begin{eqnarray}
Q_n=(\frac{c_1}{\lambda}+\frac{\tau_{n-1}^2}{\lambda^3\tau_{n-2}\tau_{n}})(\frac{\tau_{n}\tau_{n-3}}{\tau_{n-1}\tau_{n-2}}\Gamma_n-\lambda)+c_1+\frac{E^{-1}a_1}{\lambda^2}+\frac{\tau_{n-1}^2}{\lambda^2\tau_{n-2}\tau_{n}}
=\sum_{j=1}^{\infty}J_n^{(2j-1)}\lambda^{-2j},
\end{eqnarray}
where
\begin{eqnarray}
J_n^{(1)}=E^{-1}a_1+\frac{\tau_{n}\tau_{n-3}}{\tau_{n-1}\tau_{n-2}}c_1,\nonumber\\
J_n^{(2j-1)}=\frac{\tau_{n}\tau_{n-3}}{\tau_{n-2}\tau_{n-1}}w_n^{(2j-1)}c_1+\frac{\tau_{n-1}\tau_{n-3}}{\tau_{n-2}^2}w_n^{(2j-3)},\qquad j=2,3,......
\end{eqnarray}
We thus obtain the conserved densities $\rho_n^{(j)}$(j=1,2,3,....) and the associated flux $J_n^{(2j-1)}$ (j=1,2,....) for eq. (2.22), where $J_n^{(2j-1)}$ is given by equation (4.30).
\section{Conclusions}
The purpose of this article is to derive negative Volterra flows and mixed Volterra flows and their infinitely many conservation laws. By means of constructing opportune time evolution equations with negative powers of spectral parameter or with positive and negative powers of spectral parameter, and using discrete zero curvature representation, the negative Volterra flows and the mixed Volterra flows are proposed. Their Lax pairs are given. As well known, the existence of infinitely many conservation laws for the 
lattice hierarchy is very important. In the present paper, by means of the matrix Lax representation, we demonstrate the existence of infinitely many conservation laws for the proposed negative Volterra flows and mixed Volterra flows and give the corresponding conserved densities and the associated fluxes formulaically.
Thus their integrability is confirmed. Though the physical applications for the two lattice hierarchies has not been found, the property of them proposed in the paper is interesting. 
\vskip 2mm
\noindent
{\bf {Acknowledgments}}\\
The authors are indebted to Professor Xing-Biao Hu for the invaluable discussions. The project sponsored by SRF for ROCS,SEM and Hong Kong RGC grant HKBU2065/01P.
\end{document}